# Design methodology for functionally graded materials: framework for considering cracking

Zhening Yang[a], Hui Sun[a], Zi-Kui Liu[a], Allison M. Beese[a, b*]

[a] Department of Materials Science and Engineering, Pennsylvania State University, University Park, PA 16802, USA

[b] Department of Mechanical Engineering, The Pennsylvania State University, University Park, PA 16802, USA

* Corresponding author: beese@matse.psu.edu

## Abstract

In functionally graded materials (FGMs) fabricated using directed energy deposition (DED) additive manufacturing (AM), cracks may form due to interdendritic stress during solidification, the formation of deleterious phases, or the buildup of residual stresses. This study builds on our previously proposed concept of three-alloy FGM system feasibility diagrams for the identification of gradient pathways that avoid deleterious phases in FGMs by also considering solidification cracking. Here, five solidification cracking criteria were integrated into the feasibility diagrams, and equilibrium simulations were carried out based on Scheil results (termed hybrid Scheil-equilibrium simulation) to predict phase formation below the solidus temperature considering solidification microsegregation. The new feasibility diagrams were applied to four previously studied FGM systems, and the newly proposed approach predicted high crack susceptibility, detrimental phase formation, or interdendritic BCC phase formation in the experimentally observed cracking region. This demonstrates the utility of the proposed framework for crack prediction in the design of future FGMs gradient pathways.

# 1. Introduction

Functionally graded materials (FGMs) are those in which structure or composition varies with position to serve under extreme conditions, such as those common to the nuclear power generation, aerospace, and maritime sectors [1]–[4]. FGMs can be used as an approach to join dissimilar metals that cannot be welded directly, such as stainless steel 304L (SS304L) to Ti-6Al-4V, by introducing nonlinear composition paths or the addition of intermediate elements or alloys to circumvent the forming of deleterious phases, referring to brittle topologically close-packed (TCP) phases in the present work that may result in component failure during fabrication or application [5]. FGMs also enable the spatial tailoring of properties within a component; for example, a gradient region between ferritic and austenitic alloys that avoids failure normally occurring due to carbon diffusion during service [6]. The powder-based directed energy deposition (DED) additive manufacturing (AM) process has been used to fabricate FGMs through the deposition of varying mixtures of powder feedstock into the melt pool created by the laser during processing [7]. However, fabricating crack-free and robust FGMs requires avoiding compositions and processing conditions where cracks will form, which is cost-prohibitive with trial-and-error, highlighting the importance of computational tools to guide FGM design.

Previous studies have shown that FGMs made by DED AM often result in cracking [8]–[10], due to intermetallic phase formation, solidification cracking, or residual stress buildup. For example, the study of FGMs from SS304L to Ti-6Al-4V with V as an intermediate showed that the presence of Fe-Ti intermetallic phases and $\sigma$ phase resulted in cracking in the gradient regions [11]. Similarly, delamination was reported in FGMs from Ti-6Al-4V to Inconel 718

(IN718) due to intermetallic phases and a coefficient of thermal expansion (CTE) mismatch between the two terminal alloys [12], and cracks were found in Ti-6Al-4V to stainless steel FGMs due to limited miscibility between the alloys, intermetallic phase formation, and residual stress [13].

CALPHAD (CALculation of PHAse Diagrams) modeling has been used to predict the formation of potentially deleterious phases during AM, with this insight used to design composition pathways that avoid these phases. The present authors proposed the concept of feasibility maps for ternary systems to aid in avoiding regions with high deleterious phase fractions, as predicted through either equilibrium simulations or Scheil solidification simulations [5]. With this approach, a stainless steel 316L (SS316L) to Ti-6Al-4V FGM with Ni-20Cr, Cr, and V as intermediate alloys was designed, fabricated, and confirmed to contain no detrimental phases. Other researchers have developed a path-finding algorithm to identify composition pathways in high-dimensional composition spaces that avoid deleterious phase formation within FGMs based on equilibrium simulations [14]. Later, an improved path-finding algorithm was developed to add constraints to the generated path, including reducing the length of the path and considering monotonic variations in properties, namely coefficient of thermal expansion [15]. However, avoidance of brittle phase formation is not sufficient for crack-free FGMs. Cracking has been reported for FGMs in which the amounts of deleterious phases were low or zero, e.g., stainless steel to Inconel FGMs [16]–[18], SS316L to Ni-20Cr to Cr to V to Ti-6Al-4V FGMs [5], and SS304L to Ni-20Cr FGMs [19], which are due to solidification cracking or liquation cracking resulted from interdendritic precipitates or due to accumulation of thermal stress during cooling process because of CTE mismatch. Additionally, an FGM of SS304L to IN625 fabricated with DED AM but with sharp composition changes to avoid compositions known to produce cracking in ref. [20]. While no cracks were observed with low energy density, they were observed with high energy density due to the large dilution and mixing zones. This highlights the importance of considering the solidification process and crack susceptibility along the entire pathway when designing FGM compositional pathways because cracking may still occur even when trying to jump over problematic compositions due to dilution.

Solidification cracking, liquation cracking, and cracking due to thermal stress buildup can occur in FGMs fabricated by DED AM. Hot tearing can occur during fabrication, including solidification cracking due to interdendritic stress or lack of liquid backfilling during solidification shrinkage and liquation cracking due to remelting of low-melting-temperature components in previous layers [21]. The thermal stresses present during AM fabrication from the material contraction upon solidification followed by expansion and contraction with thermal cycles are exacerbated in FGMs due to mismatched coefficients of thermal expansion between phases within layers or between layers of differing compositions.

While AM processes and materials present new challenges for crack susceptibility models, their development in the context of traditional casting and welding processes is relatively well-established. For example, research on stainless steels has demonstrated the effect of solidification mode, ferrite amount, and impurities on solidification crack susceptibility. A low ferrite level (<5%) and a high ferrite level (>20%) were found to increase solidification crack susceptibility [22]. Austenite (A) and austenite-ferrite (A-F) solidification modes were found to be more susceptible to solidification cracking than ferrite-austenite (F-A) and ferrite (F) solidification modes [22]. Impurities such as P and S and alloy elements such as Ti, Nb, N were found to form compounds and eutectics, which increased solidification crack susceptibility [23]. Models to predict hot cracking in welding processes were applied to the development of Nickel alloys for additive manufacturing, including the freezing range (FR) model, where only the temperature drop during solidification is considered, the crack susceptibility coefficient (CSC) model, based on the relative time spent in crack susceptible solidification region, and the Kou criterion, where the lateral growth rate of dendrites and grains is considered [24]. The Rappaz-Drezet-Gramaud (RDG) model considers interdendritic pressure change due to solidification shrinkage and mechanical deformation, calculating a critical deformation rate to represent solidification crack susceptibility [25]. The RDG model requires material properties and solidification conditions as discussed in the next section. To obtain a simplified solidification crack susceptibility indicator based only on the solidification path, Easton et al. [26] proposed an indicator that only considered the pressure drop due to liquid shrinkage in the RDG model, here referred to as the simplified RDG or sRDG criterion, and also provided an improvement for the CSC criterion, here referred to as the improved CSC or iCSC criterion, to focus on the end of the solidification process.

In conclusion, when joining dissimilar alloys in FGMs through liquid-phase processing, new alloys are created in the gradient region, and the printability of these alloys is unknown compared to the terminal alloys. Therefore, computational methods are necessary to evaluate the feasibility of FGM gradient paths. However, existing approaches for evaluating FGM path designs only consider the presence of deleterious phases, either under the assumption of equilibrium or using the bounds of equilibrium and Scheil solidification simulations. This paper provides a framework for considering both deleterious phase formation and crack susceptibility during the solidification process in designing gradient paths for FGMs to be fabricated by AM, as well as a visualization of feasibility of three-alloy FGM systems to help design path. The FR, CSC, Kou, simplified RDG (sRDG), and improved CSC (iCSC) criteria were used to calculate solidification crack susceptibility as a function of composition, and equilibrium and Scheil simulations were used to predict phase formation. Furthermore, instead of using elemental ternary systems as a simplification of the alloy systems, ternary FGM feasibility maps with alloy compositions were calculated to aid in the visualization of deleterious phase formation and solidification crack susceptibility within multi-alloy systems. The proposed approach for evaluating compositional feasibility was applied to previously studied FGMs, including SS304L to Inconel 625 (IN625) [16], SS316L to Ni-20Cr to Cr to V to Ti-6Al-4V [5], SS304L to Ni-20Cr [19], and Ti-6Al-4V to Invar [27], comparing the computational predictions with experimentally observed compositions where cracking occurred. The micro-cracks observed in the SS304L to IN625 and SS316L to Ni-20Cr to Cr to V to Ti-6Al-4V FGMs were confirmed to be solidification cracks. The large crack in the SS304L to Ni-20Cr FGMs was attributed to the CTE mismatch between the FCC and BCC phases, while the fracture in the Ti-6Al-4V to Invar sample was determined to be primarily due to the large amount of intermetallic phases along with the CTE mismatch between the Laves C14 and BCC phases [28][29]. We also carried out equilibrium simulations using Scheil results, here termed 'hybrid Scheil-equilibrium simulation', to show how microsegregation during the AM process affected FCC-BCC phase transformation in the SS304L to Ni-20Cr FGMs, which explained the cracks in the corresponding sample. The deleterious phase feasibility map predicts regions that may cause failure during fabrication, while the solidification crack susceptibility maps predict regions that should either be avoided or require additional processing considerations to avoid cracking. For example, if a region of high crack susceptibility must be accessed, in situ heating may be added to reduce the cooling rate, or post-processing, such as hot isostatic pressing (HIP) should be used to heal cracks. Furthermore, the proposed hybrid Scheil-equilibrium simulations can provide insight into what phases can be expected during the heating caused by subsequent laser passes or post-processing heat treatment, which can aid in the design of post-process heat treatments. The framework for creating comprehensive feasibility diagrams presented herein provides a tool for designing gradient pathways for FGMs fabricated with DED AM.

2. **Experimental methods**

The following four FGMs that were previously fabricated and analyzed were examined here: SS304L to IN625 [16], SS316L to Ni-20Cr to Cr to V to Ti-6Al-4V [5], Ti-6Al-4V to Invar [27], and SS304L to Ni-20Cr [19]. All these FGMs were fabricated using DED AM by changing the volume fraction of two different powder feedstocks as a function of vertical position (see **Figure 1**). Details of the gradient paths, DED process parameters, and characterization methods are available in previous publications. In the present work, energy-dispersive X-ray spectroscopy (EDS) analysis was performed in a scanning electron microscope (SEM, Thermo-Scientific, Apreo S) with a silicon drift detector attachment (Oxford Instruments, Ultim Max silicon drift detector) to measure compositions in the crack region of the SS316L to Ni-20Cr to Cr to V to Ti-6Al-4V FGM. All other experimental data presented here are from the corresponding previously published papers.

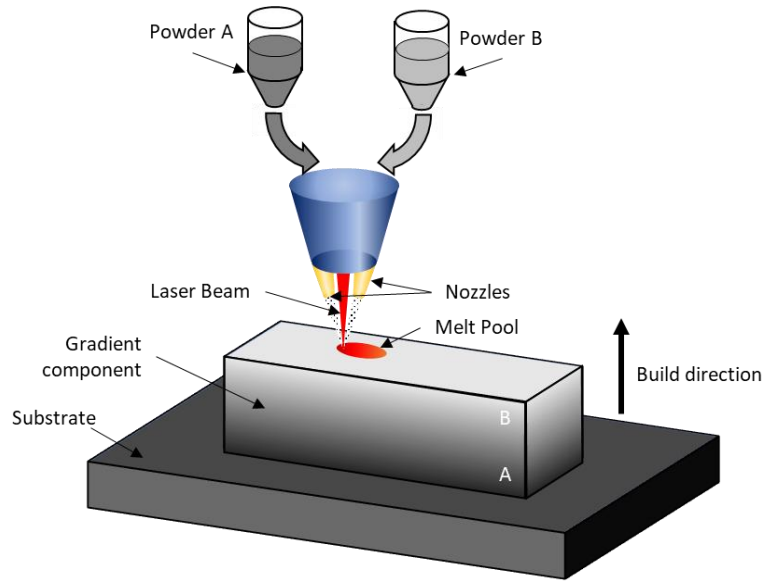

**Figure 1** Schematic of fabrication FGMs using the DED system by varying powder mixing ratio in different locations.

3. **CALPHAD methods**

The CALPHAD approach enables the simulations of multi-component multi-phase behavior in complex systems by modeling thermodynamic properties for each phase [30]. Equilibrium simulations define phases present as a function of composition and temperature at equilibrium or infinite times or infinite diffusion rates in all phases. The Scheil-Gulliver solidification model can be used to predict the amounts of phases formed during the solidification process by assuming an infinite diffusion rate and thus homogenous composition in the liquid, negligible diffusion in the solid, and local equilibrium at the solid/liquid interface [31][32]. Equilibrium and Scheil simulations based on the CALPHAD method were used in the present work to predict phase formation under two extreme conditions.

The improved feasibility maps proposed here provide a visualization of composition regions anticipated to result in deleterious phases or high solidification crack susceptibility within FGMs by merging of maps of deleterious phase predictions (via equilibrium and Scheil solidification simulations) and susceptibility of solidification cracking.

To create the deleterious phase feasibility maps, Scheil simulations were performed for each composition to predict phase formation during rapid solidification [33]. Additionally, the solidus temperature as a function of composition was extracted from Scheil simulations. Equilibrium simulations were performed over a temperature range from $\frac{2}{3}T_s$ to $T_s$, where $T_s$ is the solidus temperature for each composition. This temperature range was selected to predict phase formation during deposition, with the assumption that below $\frac{2}{3}T_s$, phase transformations are kinetically frozen out [34]. The feasibility of each composition was determined by comparing the total amount of predicted deleterious phases with a user-defined threshold, where the amount of deleterious phase in the equilibrium results was determined by summing the deleterious phase fractions at each temperature and taking the highest amount within the temperature range, while the amount of deleterious phase in Scheil simulations was taken directly from the Scheil simulation results.

The solidification process can be divided into four stages [23]: (1) nucleation, where the liquid phase is continuous, with dispersed solid nuclei; (2) the growth of solid phase while liquid can still move and fill in cracks; (3) further growth of the solid phase resulting in the disruption of liquid passages such that the dispersed liquid cannot fill voids, which is the highest solidification crack susceptibility stage; and (4) finally, the end of the solidification process when liquid phase completely disappears. Here, we evaluated composition maps in terms of five solidification crack susceptibility criteria, as shown in **Figure 2**. In each case, the solidification crack susceptibility criteria used the temperature versus solid fraction curves generated from Scheil simulations, where most of the criteria evaluated focus on the third stage of the solidification process.

The freezing range (FR) criterion, which considers the difference between the liquidus temperature ($T_L$) and solidus temperature ($T_S$), is given by:

$$FR = \Delta T = T_L - T_S. \qquad \text{(Eq. 1)}$$

Higher values of FR indicate that the composition will remain in the mushy zone over a larger temperature range, corresponding to a higher possibility to form harmful interdendritic phases than a material with a small FR.

The CSC criterion, which considers the ratio of time in the third solidification stage to that in the second solidification stage, is given by:

$$CSC = \frac{t_{f_s=X_1} - t_{f_s=X_2}}{t_{f_s=X_2} - t_{f_s=X_3}} \approx \frac{T_{f_s=X_1} - T_{f_s=X_2}}{T_{f_s=X_2} - T_{f_s=X_3}}, \qquad \text{(Eq. 2)}$$

where $t_{f_s=X_i}$ and $T_{f_s=X_i}$ are the time and temperature when the solid fraction equals Xi, and i=1,2,3, respectively. The temperature-based ratio is used here, with the values of $X_1$, $X_2$, and $X_3$ taken to be 0.99, 0.9, and 0.4, as in ref. [24]. A large value of CSC indicates a large fraction of time in the solidification process during which cracks cannot be healed by liquid filling, resulting in a high solidification crack susceptibility. In Scheil simulations, equilibrium is assumed at the solid/liquid interface, so solute atoms may be rejected from the solid and concentrate in the remaining liquid during solidification, bringing the composition of the liquid closer to the eutectic composition. A large CSC value (i.e., a large value of $T_{f_s=X_1} - T_{f_s=X_2}$) occurs when the temperature drops dramatically before the eutectic reaction and the eutectic reaction starts at the very end of the solidification process. This results in shrinkage of the matrix and a corresponding increase in the volume of the remaining liquid channels, which may turn into voids or cracks after solidification as there is no liquid remaining to backfill the channels.

In addition to a likelihood of solidification cracking, the formation of eutectic structures,

which have a much lower melting temperature than the matrix, at the very end of the solidification process, can result in liquation cracking due to reheating during subsequent laser passes during AM fabrication or post-process heat treatment. According to Eq. 2, a eutectic reaction early in the solidification process will result in a lower CSC value, i.e., a higher solidification cracking resistance. For alloys with compositions closer to the eutectic composition, if eutectic solidification occurs before $f_s = X_2$, CSC=0, indicating no susceptibility to solidification cracking. An issue with the CSC criterion is that if eutectic solidification occurs before $X_3$, the CSC value becomes undefined. In this case, we take the CSC value to be zero, indicating that the alloy at this composition is resistant to solidification cracking.

The iCSC criterion, which integrates the solid fraction as a function of temperature within the third stage of the solidification process is given as:

$$iCSC = \int_{T_{f_s=X_1}}^{T_{f_s=X_2}} f_s EqT) dT, \quad \text{(Eq. 3)}$$

where $X_1$ and $X_2$ were taken to be 0.7 and 0.98 here, the same as in ref. [26]. Compared to the CSC model, the iCSC criterion focuses on the end of the solidification process, where liquid filling to heal voids is limited and solidification crack susceptibility is the highest.

The sRDG criterion is calculated as:

$$sRDG = \int_{T_{f_s=X_1}}^{T_{f_s=X_2}} \frac{f_s(T)^2}{(1-f_s(T))^2} dT, \quad \text{(Eq. 4)}$$

where $X_1$ and $X_2$ were taken to be 0.7 and 0.98 here, the same as in ref. [26]. The sRDG criterion modifies the RDG model by considering a simplification of its term describing the pressure change due to solidification shrinkage. Unlike the RDG criterion, where materials properties (e.g., liquid viscosity, liquid and solid density) and solidification conditions (e.g., temperature gradient, secondary dendrite arm spacing) are required, Easton et al. [26] proposed the sRDG criterion, which can be calculated directly from Scheil simulations, making it more suitable for alloy development, where properties of new compositions are unknown.

The Kou criterion represents the inverse of the speed of the lateral growth of dendrites and grains at the end of the solidification process as shown below:

$$Kou = \left|\frac{dT}{d\left(f_s^{\frac{1}{2}}\right)}\right|_{f_s^{\frac{1}{2}} \to 1} \approx \left|\frac{T_{f_s=X_1} - T_{f_s=X_2}}{\sqrt{X_1} - \sqrt{X_2}}\right|, \quad (Eq.\ 5)$$

where $X_1$ and $X_2$ were taken to be 0.98 and 0.93 here, higher than the values used by Yu et al. [24] to consider the behavior near the end of solidification. A high value of the Kou criterion indicates a slow lateral growth that would result in long interdendritic with intergranular liquid channels susceptible to crack formation.

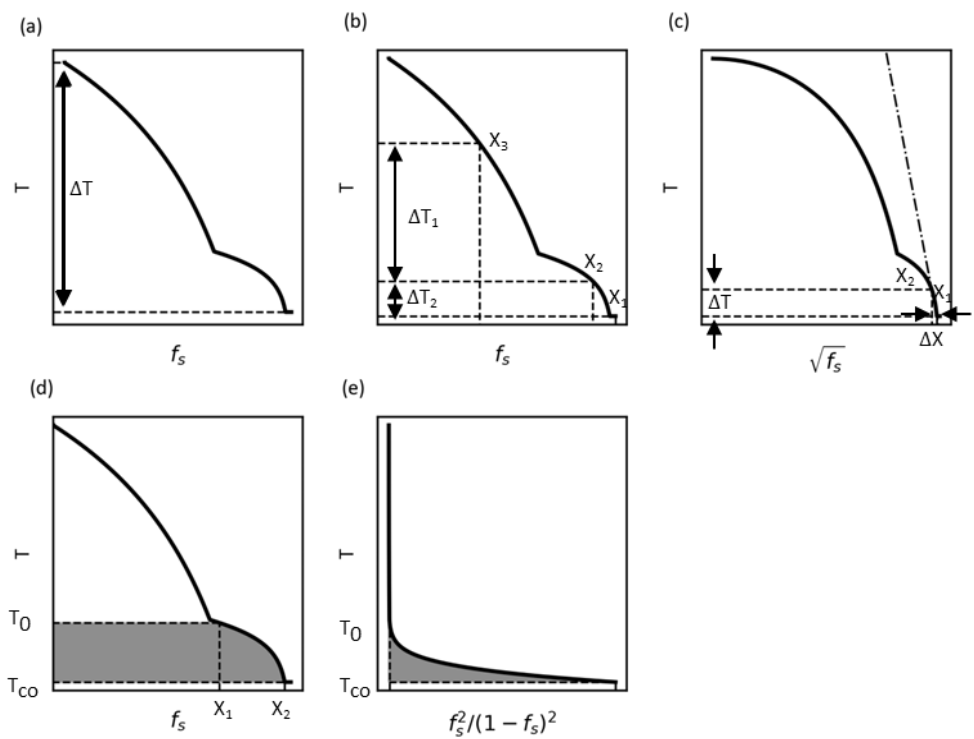

**Figure 2:** Schematics for calculating the five cracking criteria with T for temperature and $f_s$ for solid fraction: (a) FR is the temperature difference between liquidus and solidus temperature, (b) CSC is the ratio of the time for the third solidification stage over the second

solidification stage, (c) Kou is the absolute value of the slope of the T vs. $\sqrt{f_s}$ curve near the end of the solidification process, (d) iCSC is the area on the left of the T vs. $f_s$ curve during the third stage of solidification, (e) sRDG is the area on the left of the T vs. $\frac{f_s^2}{(1-f_s)^2}$ curve during the third stage of solidification, where the end of the solidification process is emphasized more than in the iCSC criterion.

For all five criteria considered here, a larger value indicates a higher likelihood of solidification cracking. In addition to solidification cracking, solid-state phase transformation during cooling in AM can introduce thermal stresses and result in cracking. However, as Scheil simulations only consider the solidification process, they cannot predict phase formation at temperatures below the solidus temperature, and equilibrium simulations use the overall composition of the layer to compute formation of phases, neglecting possible phase transformations due to microsegregation. To investigate solid-state phase transformations in interdendritic or intergranular regions during solidification, we propose to compute the phase equilibrium as a function of local, micro-segregated local compositions from Scheil simulations, i.e., the hybrid Scheil-equilibrium simulation. Phase fractions predicted in the simulations considering solidification microsegregation in the SS304L to Ni-20Cr FGM sample using the FEDEMO database from Thermo-Calc [35] are plotted in **Figure 3**.

**Figure 3a-b** uses a composition with 50 wt% Ni-20Cr/50 wt% SS304L to demonstrate the hybrid Scheil-equilibrium simulation process, while **Figure 3c** shows the results of the hybrid Scheil-equilibrium calculation at 700 °C for the SS304L to Ni-20Cr FGM by carrying out the steps shown in **(a-b)** for each composition along the gradient path. As shown in **Figure 3a** using a layer with 50 wt% Ni-20Cr/50 wt% SS304L as an example, a Scheil simulation was first performed for the layer composition. The calculated elemental composition as a function

of solid fraction is presented, providing insight on the spatial composition distribution in the corresponding as-built layer. Equilibrium simulations were then carried out for each composition along the Scheil simulation curves in **Figure 3a** with the predicted equilibrium phase fractions at 700 °C given in **Figure 3b**, which provides insight into the equilibrium phases distribution as a function of solid fraction. The area under the curve for each phase in **Figure 3b** corresponds to the overall fraction of that phase within the layer of interest. By repeating the Scheil and equilibrium simulations for each composition in the gradient region, the expected phases at 700 °C, considering microsegregation, for the SS304L to Ni-20Cr FGM are shown in **Figure 3c**. Compared to Scheil simulations and equilibrium simulations at 700 °C using the overall layer compositions, where no BCC phase formation was predicted, the results here match better the X-ray diffraction (XRD) and electron backscatter diffraction (EBSD) results from ref. [19]. The Scheil results in **Figure 3a** show an increase of Cr concentration at the end of solidification, favoring the formation of the BCC phase, which was ignored in the equilibrium simulations using layer compositions. Compared to kinetic simulation modules like DICTRA and TC-PRISMA in the commercial software Thermo-Calc [35], the hybrid Scheil-equilibrium simulation does not take diffusion into consideration. However, this proposed framework does not require mobility databases and, considering that the high cooling rate of AM processing limits the effects of diffusion, the hybrid Scheil-equilibrium simulation offers a simple way to predict amounts of phases considering solidification microsegregation in AM.

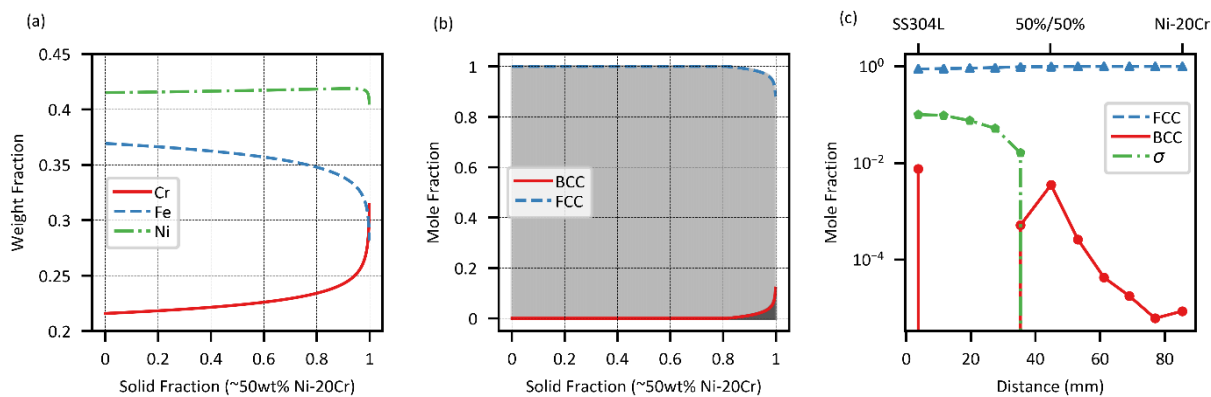

**Figure 3:** Approach for performing hybrid Scheil-equilibrium simulations. (a) Scheil-predicted composition of the solid at the solid/liquid interface as a function of solid fraction based on the overall composition of a layer, where Scheil results for a layer wise composition of 50 wt% Ni-20Cr/50% wt% SS304L is shown as an example, (b) equilibrium phase predictions at 700 °C for each composition shown in (a), where the area beneath each curve corresponds to the total mole fraction of the corresponding phase for that layer composition; here, this corresponds to ~99.65 mole% FCC and ~0.35 mole% BCC. (c) Hybrid Scheil-equilibrium calculation of phases along the SS304L to Ni-20Cr FGM at 700 °C following the procedure outlined in (a)-(b) for each layer of the FGM (EDS composition can be found in ref. [20]).

## 4. Results and Discussion

### 4.1 SS316L to Ni-20Cr to Cr to V to Ti-6Al-4V

Previous Scheil and equilibrium deleterious phase feasibility maps showed no brittle deleterious phases along the SS316L to Ni-20Cr to Cr to V to Ti-6Al-4V composition path [5]. However, micro-cracks were observed near the pure Cr region, extending into the Ni-20Cr + Cr and Cr + V regions. **Figure 4b** and **Figure 4d** show an increase in Ni concentration near the crack. Because here Ni was being pushed from solid into liquid during solidification as discussed below, i.e., the later the region solidified, the higher the Ni concentration, the increase of Ni concentration near the crack indicates that cracking occurred in the region that solidified last and was vulnerable to solidification cracking. Equilibrium and Scheil simulations were carried out for the Cr-Ni-V system using a newly development Al-Cr-Fe-Ni-Ti-V database. All five solidification crack susceptibility criteria showed a peak value near the pure Cr region in the SS316L to Ni-20Cr to Cr to V to Ti-6Al-4V FGM as shown in **Figure 5**.

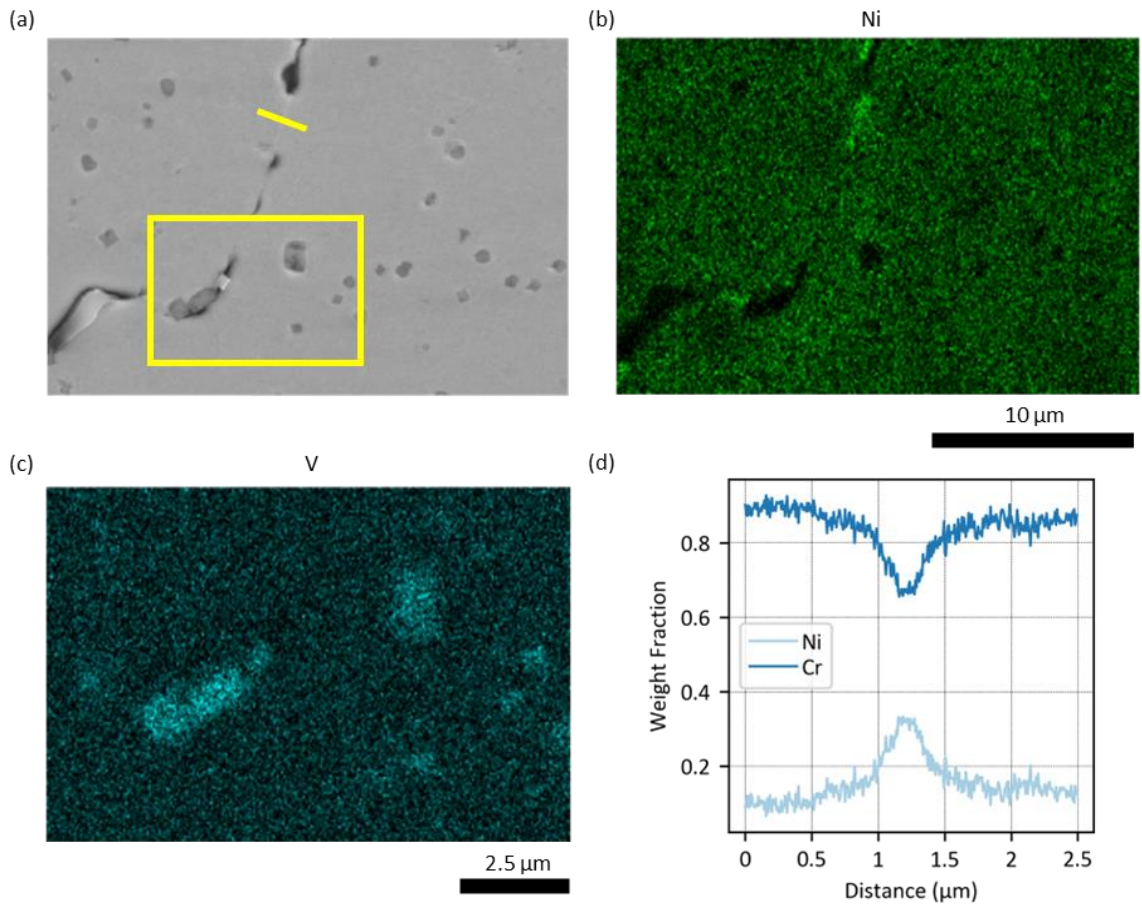

**Figure 4:** EDS maps showing the distribution of Ni and V near the cracks in the SS316L to Ni-20Cr to Cr to V to Ti-6Al-4V FGMs [5] for regions with ~87 wt% Cr, 12 wt% Ni and 1 wt% V: (a) SEM image around the crack, (b) Ni distribution map for the region in (a), (c) V distribution map for the region in yellow rectangle in (a) showing that the particles from the above layers mentioned in ref. [5] contributed to the cracking too , and (d) Ni and Cr concentrations along the yellow line in (a) shows an increase of Ni concentration near the crack, indicating the crack happens at the end of the solidification process.

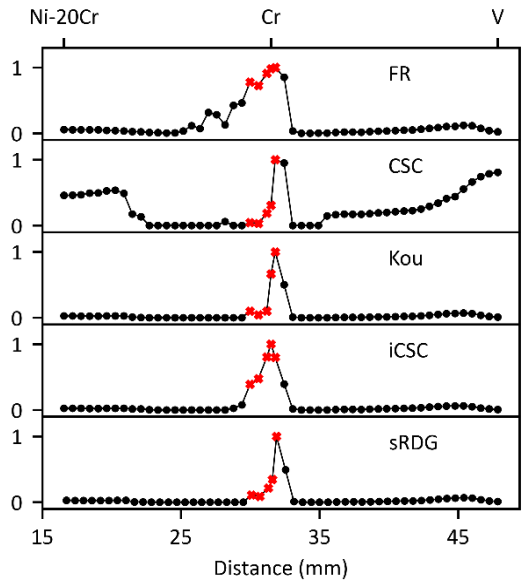

**Figure 5:** Normalized solidification crack susceptibility calculated using experimentally measured composition in the Ni-20Cr-Cr-V region of the SS316L to Ni-20Cr to Cr to V to Ti-6Al-4V FGMs (EDS composition can be found in ref. [5]), where the red "X" symbols indicate where cracks were seen experimentally (see **Figure 4**).

In the Cr-Ni binary system, the peak in the solidification crack susceptibility criteria is due to the steepness of the liquidus (i.e., large $\frac{\Delta T}{\Delta c(Ni)}$, where c(Ni) is Ni concentration) and the large solid-liquid equilibrium region (i.e., large difference between $T_S$ and $T_L$) at the Cr-rich side, indicating a drastic increase in Ni concentration in the remaining liquid with increasing solid fraction, and a large temperature drop during the solidification process, which were captured by the solidification crack susceptibility criteria. As the Ni concentration increased in the remaining liquid, its composition eventually allowed for a eutectic reaction. The further the initial composition is away from the eutectic composition (~49 wt% Ni and 51 wt% Cr), the later the eutectic reaction happens during solidification, which, as discussed in the previous section, increases solidification crack susceptibility. The actual sample is more complicated than the pure Ni-Cr system, as evidenced by the particles and the small amount of V that were found in the cracked region (shown in **Figure 4**). Adding V into the Ni-Cr system suppressed the terminal eutectic reaction, letting the temperature further drop at the end of the solidification process and increasing solidification crack susceptibility. The particles detected in the crack shown in **Figure 4a** and **Figure 4c** were reported Bobbio et al. [5] to be particles with high liquidus temperature from the above layers. It is possible that concentrated thermal stress around these particles initiated cracks during cooling.

The CSC increases in the Cr-V region as the Cr-V system has a similar liquidus curve to the Ni-Cr system. This means that microsegregation can decrease the solidus temperature that increases susceptibility of solidification cracking in this region. However, the solidification temperature range is much smaller in the Cr-V system compared to the Ni-Cr system, indicating that the Cr-V system should have a smaller solidification crack susceptibility. Among all five crack susceptibility criteria studied, only the CSC criterion uses the relative temperature ratio (Eq. 2), which oversimplifies the solidification process and overestimates the effect of microsegregation in the Cr-V system.

The Cr-Ni-V ternary system is used here as an example to demonstrate the FGM design process using the newly proposed feasibility diagrams. **Figure 6** shows the equilibrium and Scheil feasibility maps for this system, and the phases in **Figure 6b-c** depict that a direct path from Ni to V is obstructed by the $\sigma$ phase region in the middle. From the point of view of avoiding deleterious phases, the composition path should be designed near the Ni-Cr and Cr-V boundaries. However, **Figure 7** shows the solidification crack susceptibility maps for the Cr-Ni-V system, where all five criteria studied have large values near the Cr corner. If deleterious phases and solidification crack susceptibility are considered simultaneously (see **Figure 8**), all the paths from the Ni corner to the V corner are infeasible, and an intermediate element or alloy must be considered. If routing composition through regions of high crack susceptibility is necessary, the fabrication process must be optimized, for example, by adding secondary laser scans or baseplate heating to reduce the cooling rate, thereby reducing thermal stresses and microsegregation, and decreasing solidification crack susceptibility [9][36][37]..

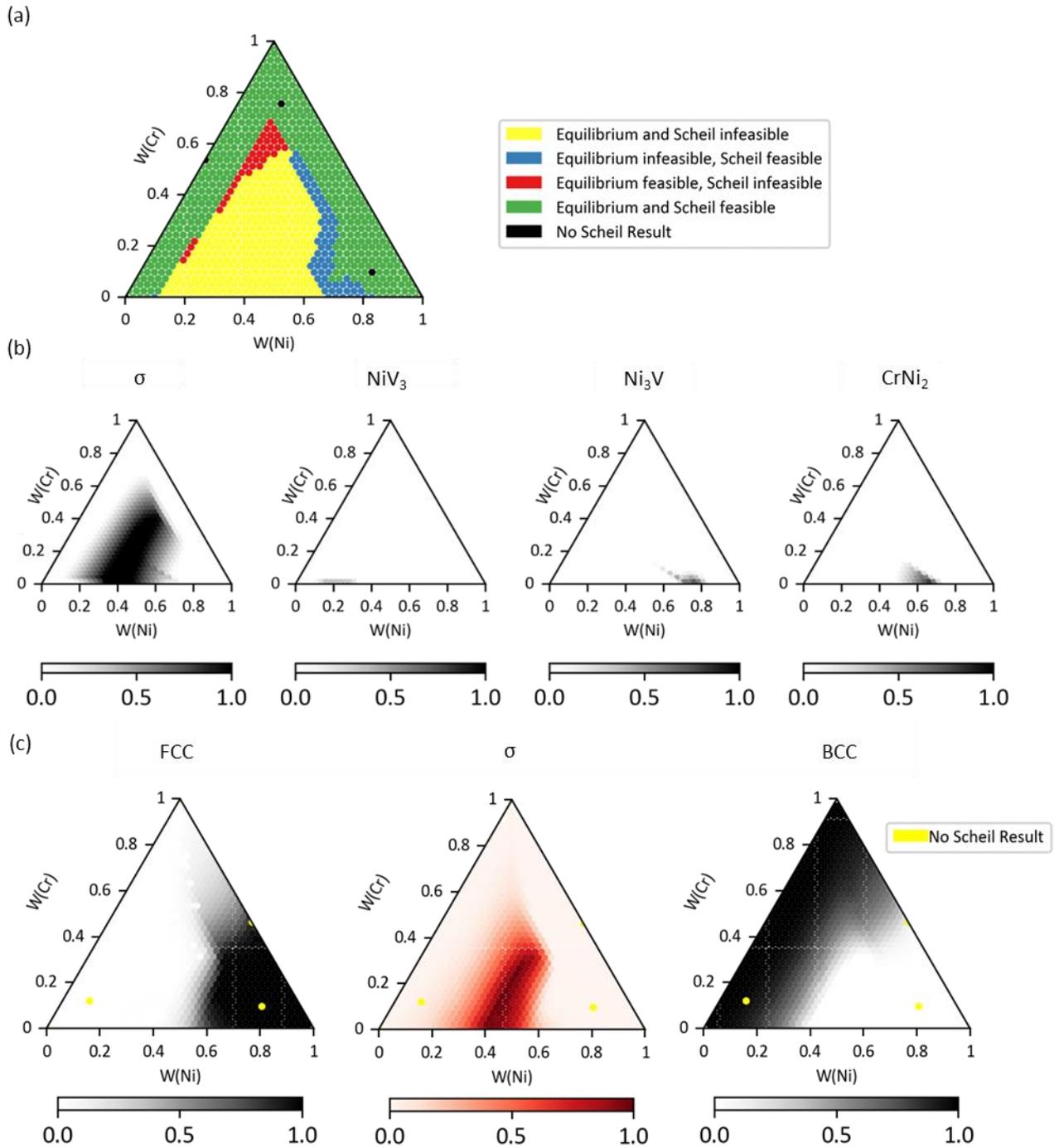

**Figure 6:** Scheil and Equilibrium simulation results for the Cr-Ni-V ternary system, where the W shown on the axis represents weight fraction. (a) Feasibility diagram considering

deleterious phase formation for the Cr-Ni-V ternary system calculated, where equilibrium simulations were performed from $\frac{2}{3}T_s$ to $T_s$, and the deleterious phase threshold was taken to be 10 mole% for equilibrium and 5 mole % for Scheil. (b) Example of equilibrium deleterious phases' distribution in the Cr-Ni-V system at 800 °C. (c) Phase distributions predicted by Scheil simulations, where σ is considered as a detrimental phase.

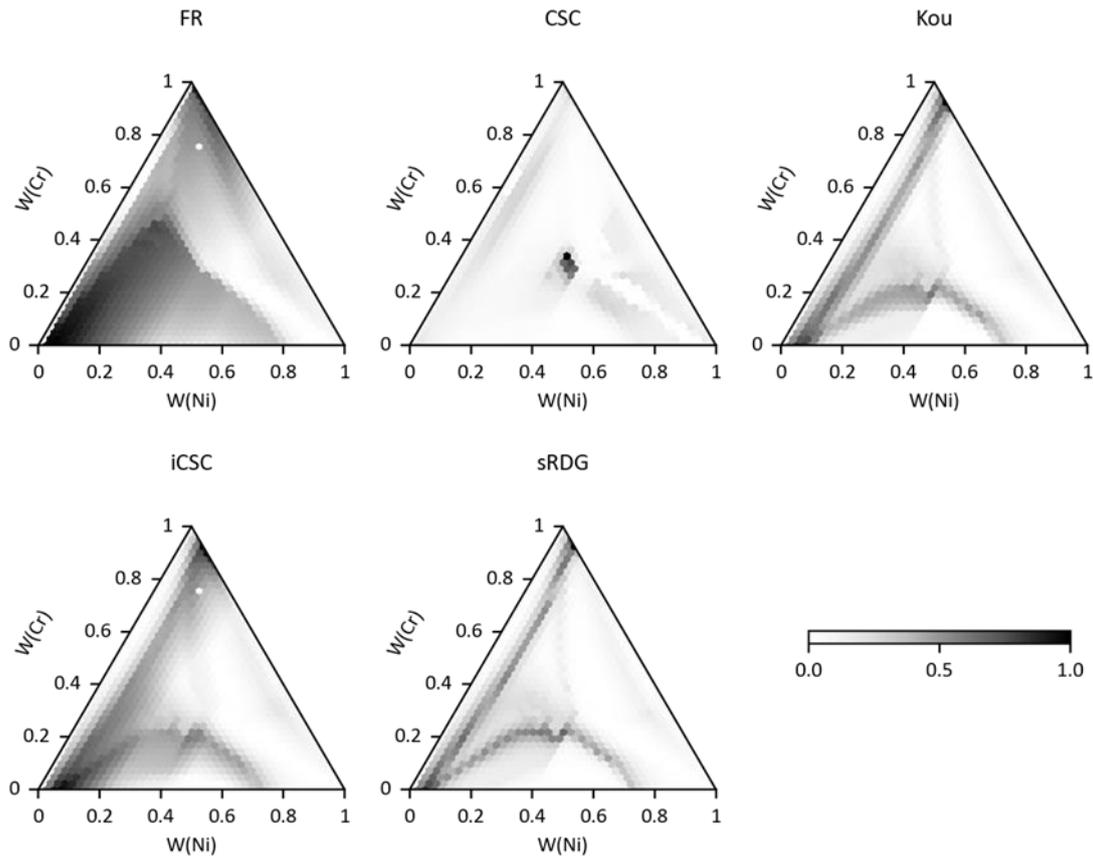

**Figure 7:** Normalized crack-susceptibility maps for the Cr-Ni-V system using the five crack-susceptibility criteria, where the W on the axis represents weight fraction.

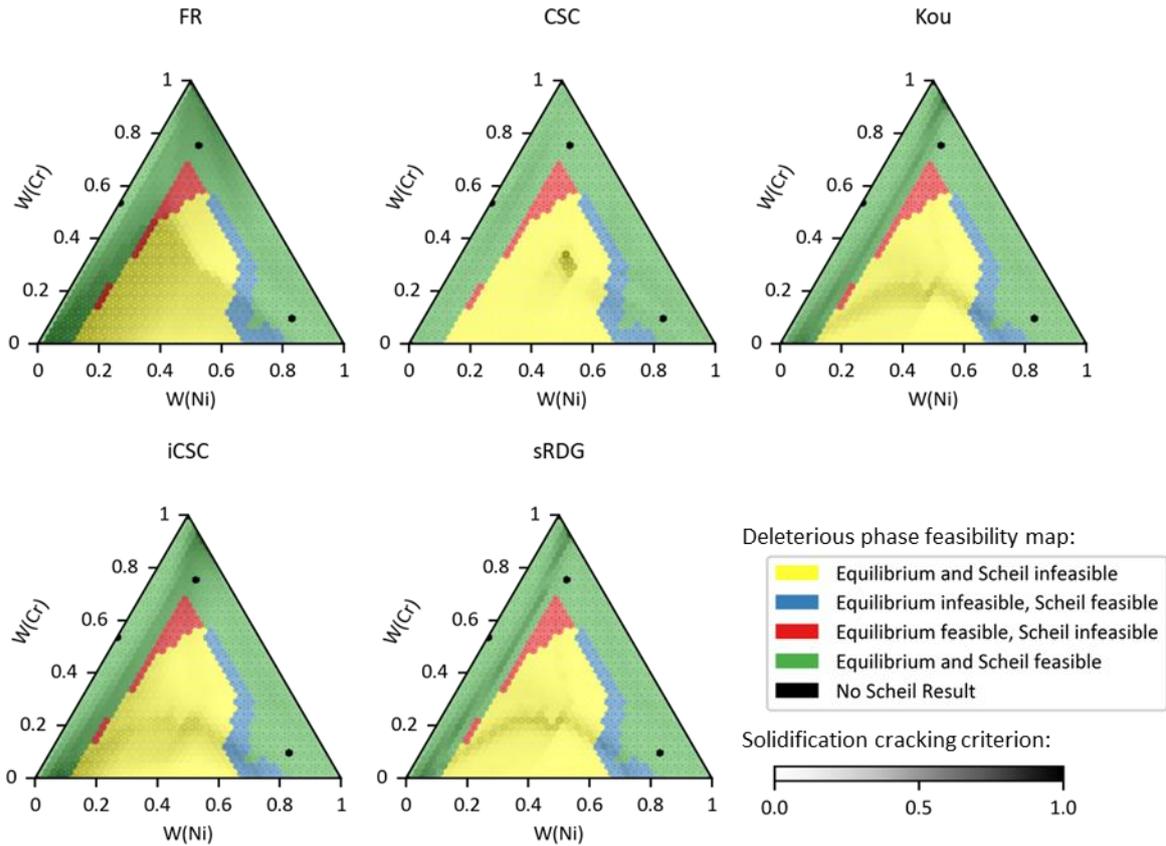

**Figure 8:** Newly proposed feasibility diagrams showing regions to avoid during DED FGM fabrication due to deleterious phases superimposed with regions to avoid due to crack susceptibility, where the W on the axis represents weight fraction.

**4.2 SS304L to Ni-20Cr to IN625**

Solidification crack susceptibility maps can also be calculated for three-alloy FGM systems. **Figure 9** shows solidification crack susceptibility maps for the SS304L-(Ni-20Cr)-IN625 system simulated with a newly developed Cr-Fe-Nb-Ni-Mo database, with compositions simplified to be the Cr-Fe-Mo-Nb-Ni system, and neglecting elements C, Mn, Si, Co, Ti, and Al due to their small amounts and database limitations. The deleterious feasibility map in **Figure 9a** shows that amounts of deleterious phases, which are Laves C14 and $\delta$ phase in this three-alloy FGM system, from both equilibrium and Scheil simulations are under the user-defined thresholds (10 mole% for equilibrium and 5 mole% for Scheil). The $\sigma$ phase was not considered in the Scheil results as it has not been fully modeled in our current thermodynamic database; however, this omission is suitable as $\sigma$ phase was neither observed in the SS304L to IN625 and SS304L to Ni-20Cr FGMs studied here, nor reported in other stainless steel to Inconel FGMs [17] and as-built IN625 parts [38], such that feasibility and solidification crack susceptibility maps should not be affected by neglecting $\sigma$ phase. The solidification crack susceptibility maps (**Figure 9b**) show that a linear gradient from SS304L to IN625 is likely to encounter solidification cracking, while if some Ni-20Cr is added in the gradient region, e.g., using a three-powder feeder system or using premixed powder, the high crack-susceptibility region can be avoided.

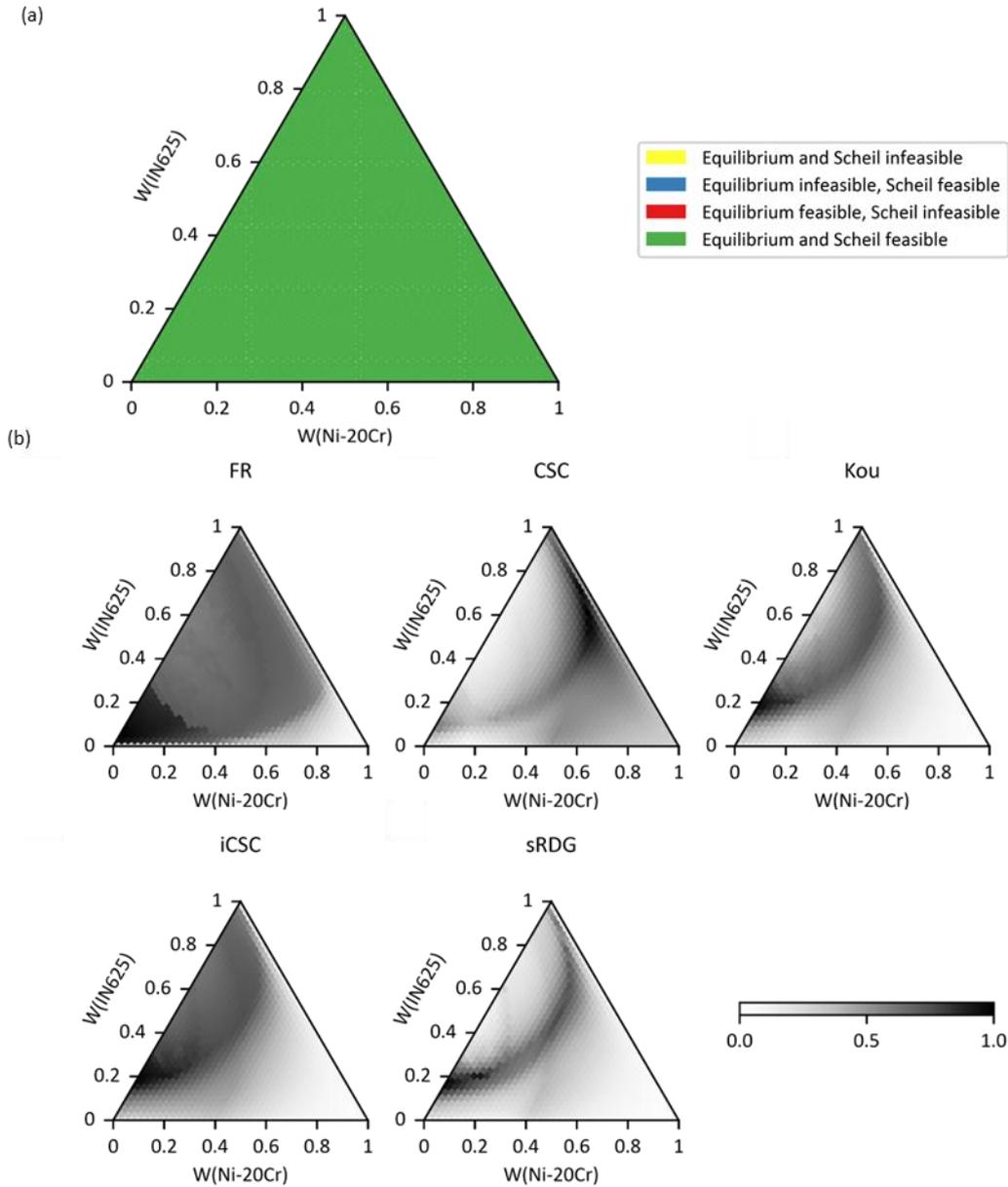

**Figure 9:** (a) Feasibility diagram considering deleterious phase formation for the SS304L-(Ni-20Cr)-IN625 system (simplified as a Cr-Fe-Mo-Nb-Ni multi-component), where equilibrium simulations were performed from $\frac{2}{3}T_s$ to $T_s$, and the deleterious phase threshold was taken to be 10 mole% for equilibrium and 5 mole% for Scheil. The W on the axis represents weight fraction. (b) Normalized solidification crack susceptibility maps based on the five criteria studied, where the W on the axis represents weight fraction.

### 4.2.1 SS304L to IN625

**Figure 10** shows the solidification crack susceptibility along the SS304L-IN625 gradient path extracted from the three-alloy FGM system shown in **Figure 9b** with the regions with experimentally observed cracking indicated. The previous analysis for this FGM showed only a trace amount of deleterious phase in the gradient region; however, cracking has been widely observed in stainless steel and Inconel joining [7], [16]–[18].

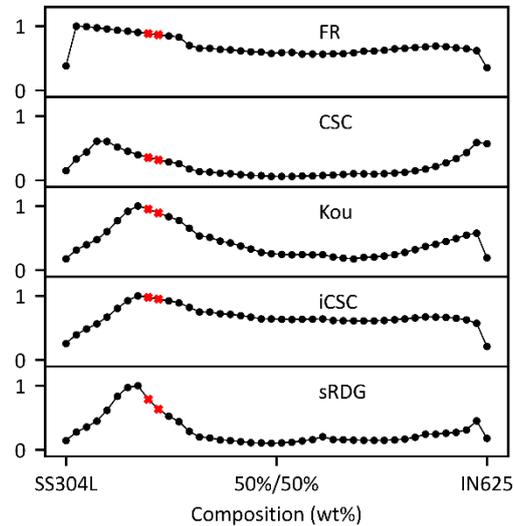

**Figure 10:** Normalized solidification crack susceptibility criteria along the SS304L-IN625 gradient pathway extracted from **Figure 9**, where the red "X" symbols indicate where cracks were observed experimentally [16].

According to the Schaeffler diagram [39] and the WRC-92 constitution diagram [40], an increase in Ni concentration reduces the ferrite fraction and pushes the solidification mode from ferrite-austenite (F-A) mode to austenite-ferrite (A-F) mode and further to austenite (A) mode. In the SS304L-IN625 FGM, a lathy ferrite phase was observed in the SS304L region, indicating an F-A solidification mode. In the SS304L-IN625 gradient region, with increased Ni from IN625, the solidification mode transformed to A-F, and the ferrite phase fraction decreased, which can affect the solidification process and increase solidification crack susceptibility [22].

**Figure 11a** shows liquidus and solidus temperatures, and **Figure 11b** shows phases from

Scheil simulations using EDS composition data for the SS304L-IN625 FGM from ref. [16], where a dramatic solidus temperature drop and increase in Laves C14 phase fraction were shown at the beginning of the gradient region. The solidification crack susceptibility criteria captured the decrease of temperature at the end of solidification, where precipitates formed, resulting in the high solidification crack susceptibility along the SS304L-IN625 boundary as shown in **Figure 9**. The experimentally observed cracking occurred with 21 wt% IN625, while the most crack susceptible region predicted in **Figure 10** corresponded to 7.3 wt% IN625 (FR and CSC criteria) and 17.1 wt% IN625 (Kou, iCSC and sRDG criteria). The reason for this discrepancy is the poor precision of Scheil simulations towards the conclusion of the solidification process, particularly when conducting simulations for multi-component systems with significant microsegregation. This is due to microsegregation potentially pushing the composition into a region that was not optimized in the database, along with the assumption of zero diffusion in the solid. The CSC criterion and Kou criterion predicted another crack susceptible region near the IN625 composition, but no cracks were experimentally observed there. Kinetic simulations in ref. [41] showed that metastable $\gamma''$ phase forms before $\delta$ phase in the interdendritic regions of IN625, which may have occurred in this sample, such that interdendritic metastable $\gamma''$ phase formed in the IN625-rich regions instead of the $\delta$ phase which Scheil simulations predicted to form. This points to a limitation of the current solidification crack susceptibility diagrams in that they do not consider metastable phases.

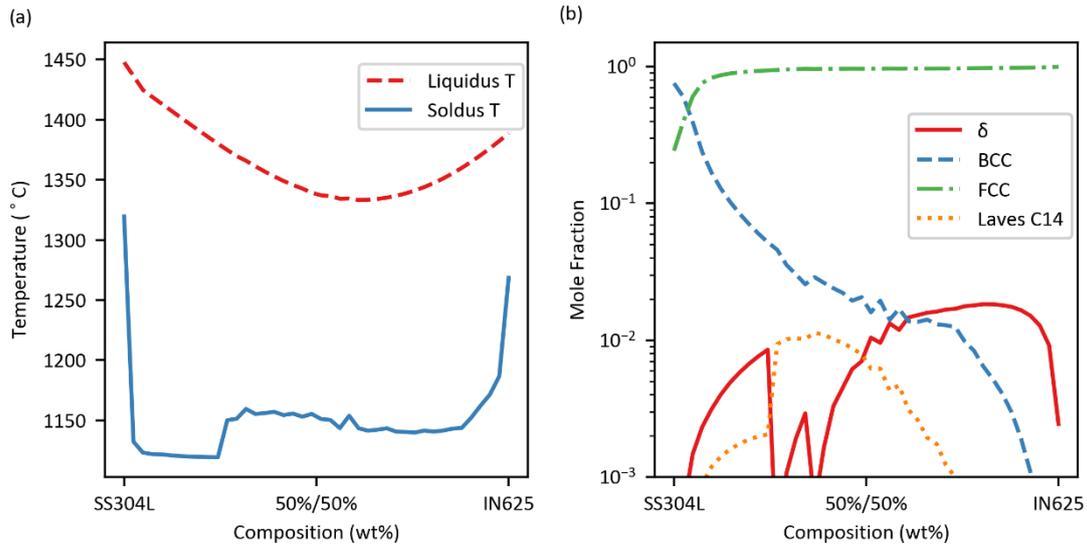

**Figure 11:** Scheil simulations for the designed composition of the linearly graded SS304L to IN625 FGM [16]: (a) solidus and liquidus temperatures showing a dramatic solidus temperature drop in the gradient region, (b) Scheil-predicted phase fractions, showing the formation of Laves C14 phase on the SS304L-rich side of the gradient region.

### 4.2.2 SS304L to Ni-20Cr

No formation of deleterious phases was predicted in the SS304L to Ni-20Cr FGM, and **Figure 12** shows the solidification crack susceptibility along the SS304L to Ni-20Cr FGM extracted from **Figure 9**, which suggests a much lower solidification crack susceptibility compared to that for the SS304L-IN625 path. While no micro-cracks were experimentally observed, a macro-scale crack was observed in the sample from the middle of the gradient region to the top of the Ni-20Cr region [19]. XRD data and EBSD maps in previous publications showed FCC phase (in Fe-rich cellular regions) and BCC phase (in Cr-rich inter-cellular regions) in the Ni-20Cr-rich gradient region, indicating that the macro-scale crack could be the result of the formation of a brittle BCC phase and thermal stresses between FCC and BCC phases during the cooling process. However, the Scheil simulation carried out in the previous publication showed only an FCC phase in the Ni-20Cr-rich region and the equilibrium simulation predicted a very low amount (< 0.1 mol%) of BCC phase formation below 600 °C. BCC phase was detected in the sample because Cr was pushed into the liquid during the solidification process (see **Figure 3a**), and the resultant higher interdendritic Cr concentration favored the formation of BCC phase. Because Scheil simulations end at the solidus temperature, and the equilibrium simulations using the overall layer composition ignores Cr microsegregation, the formation of BCC phase was underestimated in previous simulations. **Figure 13** presents a comparison between the results obtained using overall layer compositions (**Figure 13a**) and those obtained using the current hybrid Scheil-equilibrium simulations (**Figure 13b**) for the SS304L to Ni-20Cr FGM. The hybrid Scheil-equilibrium method predicted the BCC phase to form in the 50% Ni-20Cr region at 900 °C, reaching up to ~5 mole% at 600 °C, while the equilibrium simulation predicted the BCC phase to form in the same region at 650 °C, reaching up to only ~0.1 mole% at 500 °C. If this low amount of BCC were present, this is not detectable by XRD or EBSD in ref. [19]. The hybrid Scheil-equilibrium simulation predicted the BCC phase formation due to the Cr-microsegregation here better than either the Scheil or equilibrium simulations alone.

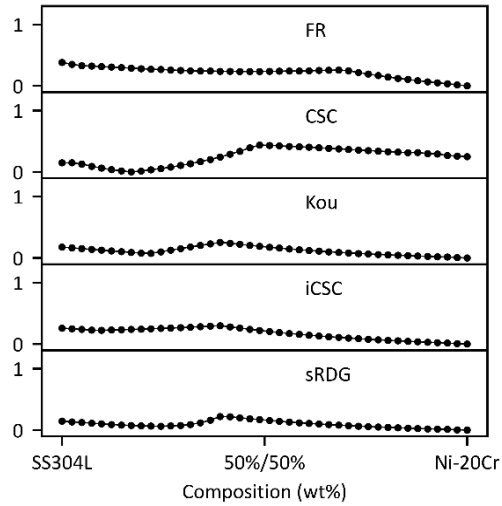

**Figure 12:** Normalized solidification crack susceptibility criteria along the SS304L to Ni-20Cr gradient pathway extracted from **Figure 9** [19], where the lack of red X marks indicates no (microscale) solidification cracking was observed experimentally.

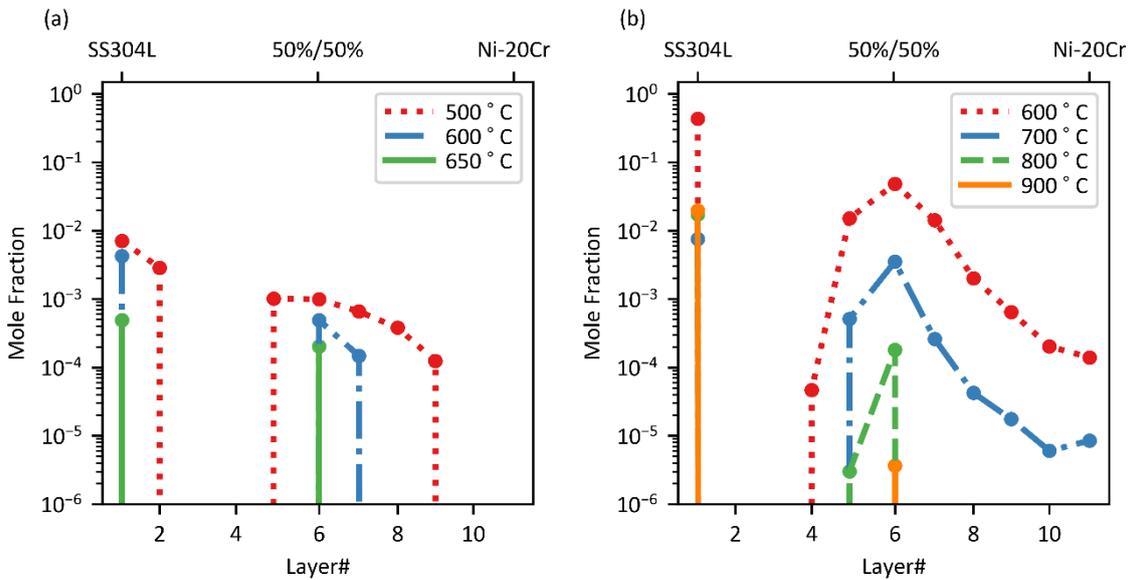

**Figure 13:** Predicted fraction of BCC phase in the SS304L-Ni-20Cr FGM at different temperatures using layer compositions calculated from EDS data (EDS data provided in ref. [19]): (a) results of equilibrium simulations using the overall layer composition, (b) results of hybrid Scheil-equilibrium simulations.

**4.3 Ti-6Al-4V to Invar**

The Ti-6Al-4V to Invar FGM sample broke into two pieces at the position with around 30-45 vol% Invar during the sectioning process after fabrication. Scheil and equilibrium simulations were carried out for this system with a Fe-Ni-Ti database developed by Keyzer et. al. [42]. High amounts of ordered BCC with about 0.5 Ti site fraction and Laves C14 intermetallic phase were predicted to form in the cracking region by Scheil simulations and were experimentally confirmed. The brittle BCC and intermetallic phases, and CTE mismatch between these phases (9.5 μm/m-°C for ordered BCC with about 0.5 Ti site fraction [28] and 8 μm/m-°C for Laves C14 phase [29]) were concluded to be the reason for cracking [27]. **Figure 14a** shows the phase fractions in the sample predicted in Scheil simulations. According to the ternary phase diagram shown in **Figure 14b**, the first BCC peak at ~8 mm distance in **Figure 14a** should be mostly disordered BCC phase with high Ti site fraction while the second BCC peak at ~12 mm distance should be mostly ordered BCC phase with high (Fe, Ni) site fraction, which is more brittle than the Ti-rich BCC [39][40].

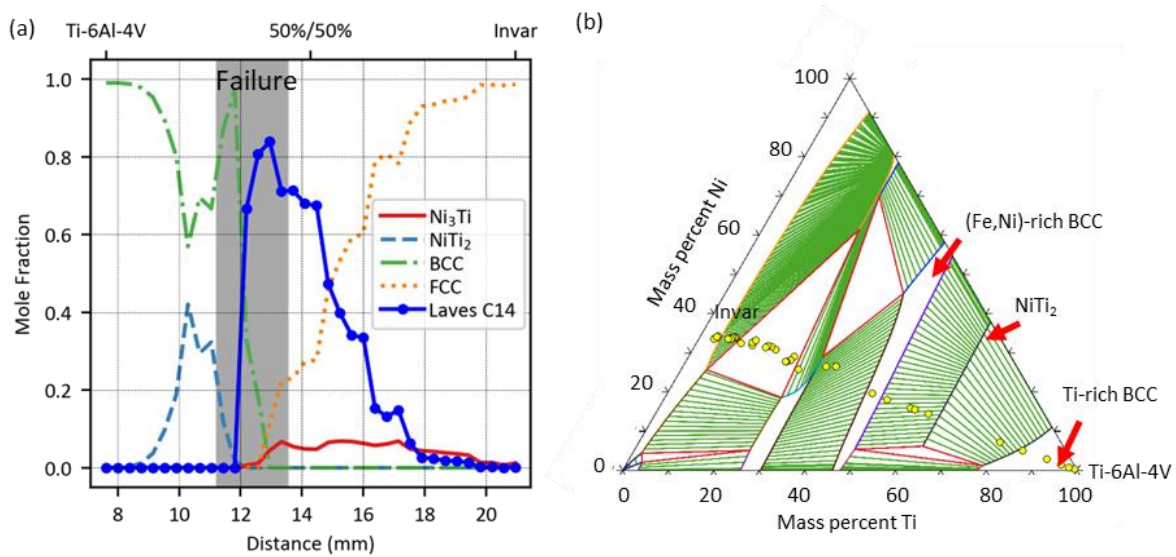

**Figure 14:** (a) Scheil-predicted phase fractions using experimentally measured composition (simplified to Fe-Ni-Ti ternary system) along the Ti-6Al-4V to Invar FGM from ref. [27], where the shaded region represents the span of experimentally observed cracking in the sample. (b) Fe-Ni-Ti ternary isothermal section at 900 °C, with the experimentally obtained EDS compositions indicated by yellow symbols.

Based on both thermodynamic simulations and experiments, a direct path from Ti-6Al-4V to Invar has already been proven to be impossible because of deleterious phase formation resulting in cracking. Solidification crack susceptibility analysis was performed for this gradient path to determine whether solidification cracking also contributed to the cracking (see **Figure 15**). All the criteria, except for the FR criterion, predicted high susceptibility in the region that cracked, indicating that solidification cracking could have contributed to the failure. High solidification crack susceptibility was also predicted in the Ti-6Al-4V to Invar gradient region near the Ti-6Al-4V composition; however, no crack was observed there since the matrix is mostly HCP phase, which have higher ductility than the Laves C14 phase or the ordered B2 phase. This highlights that neither the deleterious feasibility map nor the solidification crack susceptibility map should be used alone. The deleterious phase map checks feasibility of compositions only in terms of mechanical properties, i.e., higher fractions of deleterious phases lead to more brittle behavior that could result in material failure during fabrication. In contrast, solidification crack susceptibility maps only consider the solidification process and ignore the effect of material strength or ductility, i.e., ignoring that some phases can be more tolerant to cracking. When designing gradient pathways, paths with low deleterious phase amounts and low crack susceptibilities are desirable. Paths with high solidification crack susceptibility but within a ductile matrix (for example, FCC phase) and low deleterious phase amount are potentially feasible, but building parameters need to be carefully optimized, as mentioned in section 4.1.

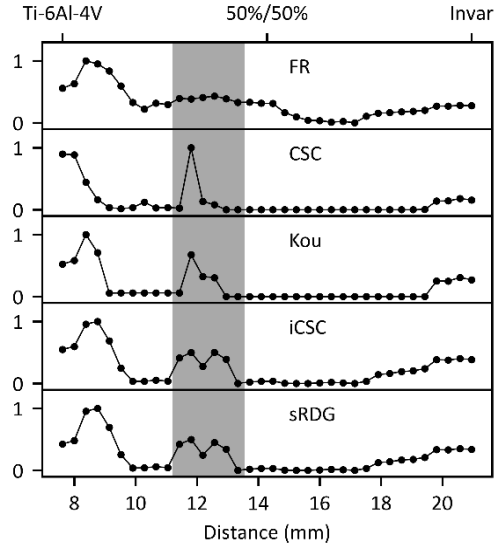

**Figure 15:** Normalized solidification crack susceptibility criteria calculated using experimentally measured composition along the Ti-6Al-4V to Invar gradient pathway from ref. [27] (simplified to Fe-Ni-Ti ternary system), where the shaded region represents the region where the sample cracked during sectioning.

## 5. Conclusions

This paper presents a comprehensive method to design gradient pathways for FGMs deposited with DED AM (or generally fusion-based processes) by taking into consideration both deleterious phase formation and the susceptibility of solidification cracking. When applying this method to evaluate four previously studied FGMs, with different terminal alloys, the newly developed feasibility maps were able to predict the compositions where cracking was experimentally observed, enabling a better understanding of the solidification process and a more robust design of FGM gradient pathways than solely considering phase formation. The main conclusions of the present study are:

- Among the five solidification crack susceptibility criteria, the Kou, iCSC and sRDG criteria accurately predicted the compositions of experimentally observed cracking, and these are thus recommended for future FGM pathway designs. Conversely, the FR criterion overestimated the cracking composition range, and as the CSC criterion underestimated the temperature decrease due to microsegregation and precipitation at the very end of the solidification process, and the predicted cracking region is shifted and sometimes incorrect.

- The proposed hybrid Scheil-equilibrium simulation can predict phase formation at temperatures below solidus temperature considering microsegregation during the solidification process, i.e., considering the interdendritic solid-state phase transformation, which is ignored in Scheil simulations of equilibrium simulations using the overall composition. Although no kinetics is considered in the simulation so the phase amount may not match the heat-treated/reheated sample, but it is still a simple way to show what phase transformation can be expected considering solidification microsegregation.

- The equilibrium and Scheil simulations successfully captured the formation of intermetallic phases and predicted failure due to their formation. The solidification crack susceptibility criteria successfully predicted solidification cracking due to solidus temperature decrease because of microsegregation and formation of precipitation. The hybrid Scheil-equilibrium method was able to predict interdendritic phase formation during the cooling process (e.g. in the SS304L to Ni-20Cr FGM), which can be used to predict stress due to CTE mismatch.

- The agreement between the newly proposed feasibility map and experiments in four different FGM systems supports the assessment of cracking possibility from Scheil and equilibrium simulations of detrimental phases, crack susceptibility criteria for solidification cracking, and hybrid Scheil-equilibrium simulations of solid-state phase transformations that supplement the deleterious phase feasibility maps. This proposed aggregate framework provides a pathway toward the design of crack-free FGMs.

**Acknowledgements:**

The author is grateful for financial support from the National Science Foundation (NSF Grant CMMI-2050069) and the Office of Naval Research (ONR Grant N00014-21-1-2608).

**Declaration of competing interest:**

None

**Data availability**

All relevant data are available from the authors.